\documentclass[aps,pra,preprint,groupedaddress,showpacs]{revtex4}
\usepackage[dvips]{graphics}
\usepackage{graphicx}
\begin{document}
%
\title{Quantum spin models with electrons in Penning traps}
\author{G. Ciaramicoli, I. Marzoli, and P. Tombesi}
\affiliation{Dipartimento di Fisica, Universit\`a degli Studi di
Camerino, 62032 Camerino, Italy}
\date{\today}
\begin{abstract}
We propose a scheme to engineer an effective spin Hamiltonian
starting from a system of electrons confined in micro-Penning
traps. By means of appropriate sequences of electromagnetic
pulses, alternated to periods of free evolution, we control the
shape and strength of the spin-spin interaction. Moreover, we can
modify the effective magnetic field experienced by the particle spin.
This procedure enables us to
reproduce notable quantum spin systems, such as Ising and $XY$
models. Thanks to its scalability, our scheme can be applied to a
fairly large number of trapped particles within the reach of near future
technology.
\end{abstract}
%
\maketitle
\section{Introduction}
Single electrons confined in Penning traps may represent a valid,
experimentally viable system for the implementation of a quantum
processor \cite{ciaramicoli1,ciaramicoli2,ciaramicoli3}. Our
proposals have been encouraged by the astonishing results obtained
in high precision experiments with a single electron
\cite{brown,peil,d'urso,d'urso2,gab} and by the advances in
trapping technology, from micro-traps \cite{drndic} to scalable
open planar Penning traps \cite{stahl,wiretrap}. In this spirit,
it has also been put forward how to realize a quantum information
channel, based on interacting spin chains, by means of trapped
electrons \cite{spinchain,gualdi}.

In this paper we focus on a linear array of electrons, each one
confined in a micro-Penning trap. Our aim is to prove that, from
the same physical system, we can derive a variety of interacting
spin models. In particular, we show how to design and control the
relevant terms in the effective spin Hamiltonian. As a result a
system of trapped electrons can be exploited to study the dynamics
of a wide range of quantum spin models. We recall that these
models are very important for the understanding of the rich
phenomenology observed in several quantum many-body systems, such
as quantum magnets and high temperature superconductors. Moreover
quantum spin systems are able to exhibit quantum phase transitions
\cite{sachdev}. To this end, it is critical to control and vary
system parameters like the applied magnetic field and the
spin-spin coupling strength. Our method to shape the effective
spin-spin interaction employs sequences of electromagnetic pulses
alternated to periods of free evolution. This technique is similar
to the refocusing schemes used in nuclear magnetic resonance (NMR)
experiments and relies, from the theoretical point of view, on the
average Hamiltonian theory \cite{nmr}. We point out that a system
of trapped electrons presents several advantages over NMR
implementations. The most important ones are scalability and the
possibility to independently adjust the values of relevant
quantities, like the spin precession frequencies and the spin-spin
coupling. Indeed their values depend on external parameters such
as the magnetic field gradient, the inter-particle distance and
the voltage applied to the trap electrodes \cite{ciaramicoli3,spinchain}.

Also other systems, like linear or planar arrays of trapped ions,
enjoy some of these properties and, therefore, it has been
proposed to use them as a quantum simulator for interacting spin
chains \cite{jane, porras1, porras2}. However working with trapped
electrons we naturally have a system of spin one-half particles,
without the need for artificially creating an effective two-level
system. Another difference between trapped ions and electrons
relies in the typical resonance frequencies. Ions are controlled
by means of a sophisticated laser setup, while trapped electrons are
manipulated by microwave or radio-frequency fields. In this
respect trapped electrons can benefit from the same technology already
developed for NMR spectroscopy.

In this paper we consider both the case of electrons with the same
spin precession frequency \cite{spinchain} and the case of
electrons with different spin precession frequencies
\cite{ciaramicoli3}. The frequency addressability, which 
is necessary to manipulate specific particles in the array, 
is obtained with the insertion of a magnetic field
gradient. 
However this condition is required only to modify the
interaction range and topology. In the case of electrons with the
same spin precession frequency, we prove that, by flipping the
spin state twice, we can effectively reduce or even cancel the
spin dynamics due to the external uniform magnetic field. This way the
effective spin system is subjected to a weaker magnetic field,
whose intensity can go down to zero, without affecting the overall
trap stability. The same resonant electromagnetic field, used to
flip the spin, is able to produce coherent superpositions of the
two spin states by adjusting its phase and duration. These
operations are the building blocks of specific pulse sequences
that allow to engineer the effective spin Hamiltonian. By
iterating such pulse sequences, we obtain various interesting spin
Hamiltonians such as the Ising model or the $XY$ model. In
addition, if we want to customize the interaction range and
coordination number, we should apply similar sequences of pulses
to selected subsets of spins in the array. The resulting spin
system can exhibit a nearest neighbor (NN) as well as a long range
interaction. The number of pulses in each sequence is relatively small and,
most notably, does not depend on the number of spins in the array,
thus making our procedure scalable.

The paper is organized as follows. In Sec. \ref{chain} we briefly
present the system of trapped electrons and review the derivation
of the effective spin-spin interaction. In Sec. \ref{extfield} we
describe how to prepare and manipulate, with an additional
oscillating magnetic field, the spin state of each electron in the
array. In Sec. \ref{varying} we show how to engineer the spin
Hamiltonian by applying appropriate electromagnetic pulse
sequences, that allow to control the strength and the range of the
interaction. The capability of our technique to reproduce a
given Hamiltonian is analyzed in Sec. \ref{fid}. Finally in Sec.
\ref{concl} we summarize our results and discuss future
perspectives. The more technical details, concerning the design of
the pulse sequences and the estimate of fidelity, are reported,
respectively, in Appendices \ref{seqs} and \ref{err}.
\section{Array of trapped electrons}
\label{chain}
Let us consider a system of $N$ electrons confined in an array of
micro-Penning traps in the presence of linear magnetic gradients.
The Hamiltonian of the system can be written as
\begin{equation}  \label{Harray}
   H = \sum_{i=1}^{N} H_i^{NC} + \sum_{i > j}^N H_{i,j}^{C},
\end{equation}
where
\begin{equation} \label{H_NC}
   H_i^{NC} = \frac{(\mathbf{p_i} -e \mathbf{A_i})^2}{2m_e} + eV_i
                    - \frac{ge\hbar}{4m_e} \mbox{\boldmath $\sigma_i$} \cdot
                    \mathbf{B_i}
\end{equation}
represents the single electron dynamics inside a trap and
\begin{equation} \label{H_C}
    H_{i,j}^{C} = \frac{e^2}{4 \pi \epsilon_0|\mathbf{r_i}-\mathbf{r_j}|}
\end{equation}
describes the Coulomb interaction between electrons $i$ and $j$.
In Eqs.~(\ref{H_NC}) and (\ref{H_C}) $m_e$, $e$, $g$, and
{\boldmath $\sigma_i$} are, respectively, the electron mass,
charge, gyromagnetic factor, and Pauli spin operators. We assume
that the micro-traps are aligned along the $x$ axis and that
$x_{i,0}$ is the position of the center of the $i$-th trap. The
electrostatic potential
\begin{equation}
    V_i(x_i, y_i, z_i) \equiv V_0 \,
                              \frac{z_i^2-[(x_i-x_{i,0})^2+y_i^2]/2}{\ell^2}
\end{equation}
is the usual quadrupole potential of a Penning trap, where $V_0$
is the applied potential difference between the trap electrodes
and $\ell$ is a characteristic trap length. The magnetic field
\cite{ciaramicoli3}
\begin{equation} \label{magfield2}
\mathbf{B_i} \equiv - \frac{b}{2} [(x_i-x_{i,0}) \mathbf{i} + y_i
\mathbf{j}]
                 + (B_{0,i} + b z_i)  \mathbf{k}
\end{equation}
is the sum of the trapping magnetic field $B_{0,i} \mathbf{k}$,
providing the radial confinement, with a local linear magnetic
gradient $b$ around the $i$-th trap.  The associated vector
potential
\begin{equation}
   \mathbf{A_i} \equiv \frac{1}{2}
          (B_{0,i} + b z_i) [ - y_i \mathbf{i} + (x_i-x_{i,0}) \mathbf{j}
          ]
\end{equation}
preserves the cylindrical symmetry of the unperturbed trapping
field.

Following the approach described in
\cite{ciaramicoli3,brown,spinchain} the Hamiltonian,
Eq.~(\ref{H_NC}), of a single electron can be written as
\begin{eqnarray} \label{H_NC_rwa}
H_i^{NC} &\simeq& -\hbar\omega_{m,i} a_{m,i}^\dagger a_{m,i}
    + \hbar\omega_{c,i} a_{c,i}^\dagger a_{c,i}
    + \hbar\omega_{z} a_{z,i}^\dagger a_{z,i}+\frac{\hbar}{2}\omega_{s,i}
    \sigma^z_{i}
    \nonumber \\
   &+& \frac{g}{4} \varepsilon \hbar \omega_z
       \left( a_{z,i}+a_{z,i}^{\dagger} \right) \sigma^z_{i}
        - \frac{g}{4} \varepsilon \hbar \omega_z
       \sqrt{\frac{\omega_z}{\omega_{c,i}}}
       \left( \sigma^{(+)}_{i} a_{c,i} + \sigma^{(-)}_{i}  a_{c,i}^\dagger
       \right)
\end{eqnarray}
where the annihilation operators $a_{m,i}$, $a_{c,i}$, $a_{z,i}$
refer, respectively, to the magnetron, cyclotron and axial
oscillators of the $i$-th electron and $\sigma^{(\pm)}_{i}\equiv
(\sigma^x_i \pm i \sigma^y_i)/2$. The frequencies of the different
electron motions are $\omega_{m,i} \simeq \omega_z^2/(2
\omega_{c,i})$, $\omega_{c,i} \simeq
(|e|B_{0,i}/m_e)-\omega_{m,i}$, $\omega_z = \sqrt{ 2e V_0/(m_e
\ell^2)}$ and $\omega_{s,i}\equiv g |e|B_{0,i}/(2 m_e)$. The
Hamiltonian (\ref {H_NC_rwa}) has been obtained under the
assumptions $\omega_{m,i} \ll \omega_z \ll \omega_{c,i}$ and
$b|z_i|/B_{0,i} \ll 1$. We also assume that the cyclotron motion
is in the ground state and the amplitude of the magnetron motion
is sufficiently small (axialization) \cite{axial}. The
dimensionless parameter
\begin{equation}
\varepsilon \equiv \frac{|e|b}{m_e \omega_z}
\sqrt{ \frac{\hbar}{2 m_e \omega_z}}
\end{equation}
represents the coupling, due to the
magnetic gradient, between internal and external degrees of
freedom of the particle.

Similarly, if the oscillation amplitude of the electrons is much
smaller than the inter-trap distance, the part of the Hamiltonian
describing the Coulomb interaction can be written as
\cite{spinchain}
\begin{eqnarray}
\label{HC5}
  H^{C}_{i,j} & \simeq &  \hbar \xi_{i,j}
  (a_{z,i}+a^{\dagger}_{z,i})(a_{z,j}+a^{\dagger}_{z,j})
- \hbar \xi_{i,j} \frac{\omega_z}{\omega_{c,i}} \left( a_{c,i}
a_{c,j}^\dagger
                                +a_{c,i}^\dagger a_{c,j}
                          \right),
\end{eqnarray}
where $\xi_{i,j} \equiv e^2/(8 \pi \epsilon_0 m_e \omega_z
         d_{i,j}^3)$ with $d_{i,j}$ being the
         distance between the $i$-th and $j$-th particle.
Now we apply to the system Hamiltonian the unitary transformation
\cite{wunderlich}
\begin{eqnarray} \label{S}
S &=& \sum_{i=1}^N \frac{g}{4} \varepsilon
  \left[\sigma^z_{i} (a_{z,i}^\dagger-a_{z,i})
        +\frac{\omega_z}{\omega_{a,i}}
         \sqrt{\frac{\omega_z}{\omega_{c,i}}}
        \left( \sigma^{(-)}_{i} a_{c,i}^\dagger -\sigma^{(+)}_{i} a_{c,i}
        \right)
  \right] ,
\end{eqnarray}
with $\omega_{a,i}\equiv\omega_{s,i}-\omega_{c,i}$. This
transformation formally removes, to the first order in
$\varepsilon$, the interaction between the internal and the
external degrees of freedom in Hamiltonian (\ref{H_NC_rwa}) and,
at the same time, introduces a coupling between the spin motions
of different electrons. Consequently the spin part of the system
Hamiltonian can be recast as \cite{spinchain}
\begin{eqnarray} \label{Hchain}
H_s  \simeq  \sum_{i=1}^{N}\frac{\hbar}{2}\omega_{s,i} \sigma^z_i
+ \frac{\hbar}{2} \sum_{i>j}^N
        \left( 2   J^{z}_{i,j} \sigma^z_i \sigma^z_j
             - J^{xy}_{i,j} \sigma^x_i \sigma^x_j
             - J^{xy}_{i,j} \sigma^y_i \sigma^y_j
        \right),
\end{eqnarray}
where
\begin{eqnarray} \label{Jz}
   J^{z}_{i,j} &=& \left(\frac{g}{2}\right)^2 \xi_{i,j}\varepsilon^2 =
   \left(\frac{g}{2}\right)^2\frac{\hbar e^4}{16\pi \varepsilon_0 m_e^4} \ \frac{b^2}{\omega_z^4 d^3_{i,j}}, \\
   J^{xy}_{i,j} &=& \left(\frac{g}{2}\right)^2 \xi_{i,j} \varepsilon^2
                    \frac{\omega_z^4}{4\omega_{a,i}^2 \omega_{c,i}^2}=
                    \left(\frac{g}{2}\right)^2\frac{\hbar e^4}{64\pi \varepsilon_0 m_e^4} \ \frac{b^2}{\omega_{a,i}^2 \omega_{c,i}^2 d^3_{i,j}}.
   \label{Jxy}
\end{eqnarray}
The effective spin Hamiltonian (\ref{Hchain}) exhibits a long
range interaction between all the particles in the chain. The
coupling strength decreases with the third power of the distance
between particles, i.e. with a dipole-like behavior. Moreover,
$J^z_{i,j}$ and $J^{xy}_{i,j}$ depend, respectively, on the axial
frequency and the cyclotron and anomaly frequencies. Since the
trapping frequencies form a well defined hierarchy, the coupling
in the longitudinal and transverse direction can be utterly
different. For example, for typical experimental values of the
cyclotron and axial frequencies, such as $\omega_c/2 \pi\simeq100$ GHz
and $\omega_z/2 \pi\simeq100$ MHz, the ratio $J^{xy}_{i,j}/J^z_{i,j}$
is less than $10^{-6}$. Therefore, for practical purposes
$J^{xy}_{i,j}$ is often negligible with respect to $J^z_{i,j}$. In
particular, this is true when the difference between the spin
frequencies of different particles is much larger than their $xy$
spin-spin coupling strength. In this case the spin Hamiltonian
reduces to
\begin{equation}\label{Hnmr}
H_s  \simeq  \sum_{i=1}^{N}\frac{\hbar}{2}\omega_{s,i} \sigma^z_i
+ \hbar \sum_{i>j}^N
            J^{z}_{i,j} \sigma^z_i \sigma^z_j.
\end{equation}
In Hamiltonian Eq. (\ref{Hnmr}) we have used the rotating wave
approximation (RWA) to neglect the interactions between spins
along the $x$ and $y$ directions, since they give rapidly rotating
terms. The Hamiltonian (\ref{Hnmr}) is, therefore, similar to the
nuclear spin Hamiltonian of the molecules used to perform NMR
experiments \cite{nmr}. However in NMR systems the spin frequency
differentiation and the spin-spin couplings are determined by the
chemical nature of the molecules, whereas in our system they
depend on the value of the applied fields, that are under control
of the experimenter.

\section{Spin state manipulation}
\label{extfield}

In this section we describe how to prepare and manipulate the spin
state with an external oscillatory field. Let us consider a
magnetic field $\mathbf{b}_p(t)$ oscillating in the $xy$ plane
with frequency $\omega$ and phase $\theta$ such that
\begin{equation} \label{bp}
\mathbf{b}_p(t) = b_p [\mathbf{i}\cos(\omega
t+\theta)+\mathbf{j}\sin(\omega t+\theta)].
\end{equation}
If we add this field to the system, the spin Hamiltonian, Eq.
(\ref{Hchain}), becomes (here and in the rest of the paper we set
$\hbar=1$)
\begin{equation}\label{Hp}
H\simeq \frac{1}{2}\sum_{j=1}^N  \omega_{s,j}
\sigma_j^z+\frac{\chi}{2}\sum_{j=1}^N [\sigma^{(+)}_{j}
e^{-i(\omega t+\theta)}+\sigma^{(-)}_{j} e^{i(\omega t+\theta)}],
\end{equation}
with $\chi \equiv g |e|b_p/(2m_e)$. In deriving the Hamiltonian
(\ref{Hp}), we assumed that the interaction between the electrons
and the oscillating magnetic field is much stronger than the
spin-spin coupling. Hence, the terms in Eq. (\ref{Hchain})
proportional to $J_{i,j}^z$ and $J_{i,j}^{xy}$ can be
neglected. In the case of a system with spin frequency
differentiation the field (\ref{bp}), applied for an appropriate
time $t$ with frequency $\omega=\omega_{s,j}$, affects only the
spin states of the resonant $j$-th electron
\begin{eqnarray} \label{puev1}
| \downarrow \rangle_j  &\rightarrow& e^{i(\omega_{s,j}/2)t} \cos
\left(\frac{\chi t}{2}\right)|\downarrow \rangle_j - i
e^{-i(\omega_{s,j}/2) t-i\theta} \sin \left(\frac{\chi
t}{2}\right)|\uparrow \rangle_j ,
\\ \label{puev2}
| \uparrow \rangle_j  &\rightarrow& e^{-i(\omega_{s,j}/2)
t}\cos\left(\frac{\chi t}{2}\right)|\uparrow \rangle_j  -i e^{i
(\omega_{s,j}/2) t+i\theta} \sin \left(\frac{\chi
t}{2}\right)|\downarrow \rangle_j.
\end{eqnarray}
Without spin frequency differentiation, the single qubit
addressing with microwave radiation is, of course, no longer
possible. Therefore when all the spins have the same precession
frequency $\omega_s$, a single resonant pulse suffices to produce
the evolution of Eqs. (\ref{puev1}) and (\ref{puev2}) for each
particle in the array. From Eqs. (\ref{puev1}) and (\ref{puev2})
we see that by changing duration and phase of the applied pulse we
can prepare and manipulate at will the spin states of the trapped
electrons. In particular, if we apply a pulse for a time
$\bar{t}=\pi/\chi$ with $\theta=0$, we can flip the spin state of
each particle
\begin{eqnarray} \label{flip1}
| \downarrow \rangle_j  &\rightarrow& -i
e^{-i(\omega_{s,j}/2)\bar{t}} |\uparrow \rangle_j , \\
\label{flip2}
| \uparrow \rangle_j  &\rightarrow& -i
e^{i(\omega_{s,j}/2)\bar{t}}|\downarrow \rangle_j.
\end{eqnarray}
We define this transformation as
\begin{equation} \label{F}
F\equiv \bigotimes_{j=1}^{
N}\{-i[\sigma^{(+)}_{j}e^{-i(\omega_{s,j}/2)\bar{t}}+\sigma^{(-)}_{j}e^{i(\omega_{s,j}/2)\bar{t}}]\}.
\end{equation}
It is not difficult to verify that the inverse transformation
$F^{-1}$ is obtained with a pulse of the same duration $\bar{t}$
but with phase $\pi$.

If we move to the interaction picture (IP) with respect to the
Hamiltonian $\sum_{i=1}^N (\omega_{s,i}/2) \sigma^z_i$, the system
evolution is given by Eqs. (\ref{puev1}) and (\ref{puev2}) with
$\omega_{s,j}=0$. Consequently, the spin flip operation, Eqs.
(\ref{flip1}) and (\ref{flip2}), turns into
\begin{eqnarray}
| \downarrow \rangle_j &\rightarrow& -i |\uparrow
\rangle_j, \\
| \uparrow \rangle_j &\rightarrow& -i |\downarrow \rangle_j.
\end{eqnarray}
 The above transformations correspond to the application of the operator
 $-i\sigma_j^x$. In a similar way a pulse applied for the time $\bar{t}$
 with $\theta=\pi/2$ produces a transformation corresponding to the
application of $-i\sigma_j^y$. Furthermore, always working in IP,
if the pulse is applied for a time $\bar{t}/2$, we can obtain the
pseudo-Hadamard operations
\begin{eqnarray}\label{p_x}
G_x &\equiv& \bigotimes_{j=1}^N \frac{(\openone-i\sigma_j^x)}{\sqrt{2}} \qquad  \text{for} \ \theta=0, \\
\label{pxdag}
G_x^{\dagger} &\equiv& \bigotimes_{j=1}^N \frac{(\openone+i\sigma_j^x)}{\sqrt{2}} \qquad  \text{for} \ \theta=\pi,\\
\label{p_y}
G_y &\equiv& \bigotimes_{j=1}^N \frac{(\openone-i\sigma_j^y)}{\sqrt{2}} \qquad \text{for} \ \theta=\frac{\pi}{2},\\
\label{pydag}
G_y^{\dagger}&\equiv& \bigotimes_{j=1}^N
\frac{(\openone+i\sigma_j^y)}{\sqrt{2}} \qquad \text{for} \
\theta=-\frac{\pi}{2}.
\end{eqnarray}
The coherent superposition of the spin states $|\uparrow\rangle$,
$|\downarrow\rangle$ for each particle can be achieved with a
single multi-frequency pulse. Hence, an appropriate choice of the
frequency, duration and phase of the pulses allows for performing,
apart from irrelevant phase factors, single qubit operations on
each spin of the array.

\section{Engineering the spin Hamiltonian} \label{varying}

In this section we show that, by using the additional magnetic
field (\ref{bp}), we can also adjust and control the form of the
effective spin Hamiltonian, starting from the models given by Eqs.
(\ref{Hchain}) and (\ref{Hnmr}). This is achieved by applying to
the system specific sequences of pulses alternated to periods of
free evolution. Our approach is inspired to the refocusing
schemes used in NMR experiments \cite{nmr}. Similarly to this technique a key point is
the choice of the different time scales.
Spin operations, operated by means of pulses, should be virtually instantaneous with respect to the free evolution of the system.
Therefore, the pulse duration should be much shorter than the free evolution time.
\subsection{Tuning of the effective magnetic field}
The spin Hamiltonian, Eq. (\ref{Hchain}), in the case of spins
with the same precession frequency can be recast as
\begin{eqnarray}\label{Heff1}
H_s &\simeq& H_0+H_c,
\end{eqnarray}
where
\begin{eqnarray}
H_0 &\equiv& \sum_{i=1}^N\frac{\omega_{s}}{2} \sigma^z_i, \\
H_c &\equiv& \frac{1}{2} \sum_{i>j}^N
        \left( 2   J^{z}_{i,j} \sigma^z_i \sigma^z_j
             - J^{xy}_{i,j} \sigma^x_i \sigma^x_j
             - J^{xy}_{i,j} \sigma^y_i \sigma^y_j
        \right).
\end{eqnarray}
In the following we shall prove that, by sending resonant pulses
of the kind of Eq. (\ref{bp}), it is possible to reduce or even
cancel the effects on the spin dynamics of the Hamiltonian term
$H_0$. This result corresponds to an effective modulation of the
external magnetic field, without affecting the trapping stability
of the whole set up.

In particular, by applying a sequence consisting of a pulse
producing the spin flip transformation $F$, Eq. (\ref{F}),
followed by a period of free evolution $t$ and by a pulse
producing the inverse transformation $F^{-1}$, we can change the
sign of the Hamiltonian term $H_0$
\begin{equation}\label{Bred}
F^{-1}e^{-iH_s t} F  = \exp[-i(-H_0 +H_c)t].
\end{equation}
To prove Eq. (\ref{Bred}) we use the identity
\begin{equation}
F^{-1}e^{-iH_s t} F=\exp[-i(F^{-1}H_0F+ F^{-1}H_cF)t].
\end{equation}
Now we have
\begin{eqnarray}
F^{-1}H_0F &=& \sum_{j=1}^N
[\sigma^{(+)}_{j}e^{-i(\omega_{s}/2)\bar{t}}+\sigma^{(-)}_{j}e^{i(\omega_{s}/2)\bar{t}}]\left(\frac{\omega_{s}}{2}
\sigma^z_j\right)[\sigma^{(+)}_{j}e^{-i(\omega_{s}/2)\bar{t}}+\sigma^{(-)}_{j}e^{i(\omega_{s}/2)\bar{t}}]
\nonumber \\
&=&-\sum_{j=1}^N\frac{\omega_{s}}{2} \sigma^z_j=-H_0.
\end{eqnarray}
The identity $F^{-1}H_cF=H_c$ follows from the commutation
relation $[H_c,F]=0$, which can be verified with some algebra.
Moreover we observe that $[H_0,H_c]=0$, because the interaction
Hamiltonian preserves the total magnetization $\sum_{i=1}^N
\sigma^z_i$. From this last consideration and from Eq.
(\ref{Bred}) we find
\begin{equation} \label{eff2}
F^{-1}e^{-iH_{s} t_2} F e^{-i H_s t_1} =\exp[-i H_{\text{eff}}
(t_1+t_2)],
\end{equation}
with
\begin{equation}
H_{\text{eff}} \equiv \frac{t_1-t_2}{t_1+t_2}\, H_0 +H_c.
\end{equation}
The left hand side of relation (\ref{eff2}) represents a sequence
consisting of a period $t_1$ of free evolution, a pulse producing
the transformation $F$, a period $t_2$ of free evolution and a
pulse producing $F^{-1}$. From the right hand side of Eq.
(\ref{eff2}), we see that this sequence is equivalent to the
system evolution for the total time $t_1+t_2$ according to the
Hamiltonian $H_{\text{eff}}$. Hence, we can obtain an effective
reduction, by a factor $(t_1-t_2)/(t_1+t_2)$, of the Hamiltonian
term $H_0$. This result can be viewed as a decrease of the
magnitude of the uniform magnetic field as far as the electron
spin dynamics is concerned. Notice that for $t_1=t_2$ we can
completely suppress the dynamical effects due to the term $H_0$.

\subsection{Design and control of the spin-spin coupling}
Let us now consider a system with spin frequency differentiation.
If we add another field consisting of a superposition of
terms resonant with the spin frequencies
\begin{equation} \label{bs}
\mathbf{b}_s(t) = \sum_{k=1}^N b_s[\mathbf{i}\cos(\omega_{s,k}
t)+\mathbf{j}\sin(\omega_{s,k} t)],
\end{equation}
the spin Hamiltonian of Eq. (\ref{Hnmr}) becomes in IP with
respect to $\sum_{i=1}^N (\omega_{s,i}/2) \sigma^z_i$
\begin{equation}
H^{IP}\simeq H^z+H^{bs},
\end{equation}
with
\begin{eqnarray}\label{Hzeta}
H^z &\equiv& \sum_{i>j}^N J^z_{i,j} \sigma^z_i \sigma^z_j, \\
H^{bs}&\equiv& \eta
\sum_{i=1}^N\left(\sigma^{(+)}_i+\sigma^{(-)}_i \right)=\eta
\sum_{i=1}^N \sigma^x_i,\label{H_bs}
\end{eqnarray}
where $\eta\equiv g|e|b_s/(4m_e)$. Hence, the application of the
oscillating field (\ref{bs}) gives rise to an effective
static transverse magnetic field, whose strength can be controlled
and modified, since it depends on the field amplitude $b_s$. This
tool may turn out useful in reproducing quantum models like Ising
system of spins. In this case the parameter $\eta$ should be
comparable to the coupling strength $J^z_{i,j}$ between the spins.

Moreover, we can engineer the spin-spin coupling, that is
introduce an effective spin-spin interaction along the $x$ and $y$
axes. This is achieved by means of sequences of pulses, of the
kind given in Eqs. (\ref{p_x}), (\ref{pxdag}), (\ref{p_y}), and
(\ref{pydag}), affecting all the spins in the array. Indeed, it
can be easily proved that
\begin{eqnarray} \label{px}
G_x e^{-iH^zt}G_x^{\dagger}&=& e^{-iG_xH^zG_x^{\dagger}t}=\exp\left(-i \sum_{j>k}^N J^z_{j,k} \sigma^y_j \sigma^y_k t \right), \\
G_y e^{-iH^zt}G_y^{\dagger}&=& e^{-iG_yH^zG_y^{\dagger}t}=\exp
\left(-i\sum_{j>k}^N J^z_{j,k} \sigma^x_j \sigma^x_k t \right)
\label{py}.
\end{eqnarray}
Hence a sequence of two specific pulses, alternated to a period of
free evolution under the Hamiltonian $H^z$, effectively modifies
the direction of the spin-spin coupling. Now, if we combine the
three operations (\ref{H_bs}), (\ref{px}), and (\ref{py}) we have
that for $t_1, t_2, t_3 \ll \pi/J^z_{i,i+1}, \pi/\eta$
\begin{equation}\label{sHxyz}
e^{-i(H^z+H^{bs})t_1}(G_y e^{-iH^zt_2}G_y^{\dagger})(G_x
e^{-iH^zt_3}G_x^{\dagger}) \simeq e^{-iH_{\text{eff}}
(t_1+t_2+t_3)},
\end{equation}
with
\begin{equation}\label{Hxyz}
H_{\text{eff}} =\tau_1 \eta \sum_{i=1}^N \sigma^x_i + \tau_1
\sum_{i>j}^N J^z_{i,j} \sigma^z_i \sigma^z_j + \tau_2 \sum_{i>j}^N
J^z_{i,j} \sigma^x_i \sigma^x_j +\tau_3 \sum_{i>j}^N J^z_{i,j}
\sigma^y_i \sigma^y_j,
\end{equation}
where $\tau_i=t_i/(t_1+t_2+t_3)$. In deriving relation
(\ref{sHxyz}) we used the approximate identity \cite{chuang}
\begin{equation} \label{tro}
e^{-iA_1 t_1}e^{-iA_2 t_2}\dots e^{-iA_n t_n} \simeq e^{-i(\tau_1
A_1+\tau_2 A_2+\dots+\tau_n A_n)t}
\end{equation}
with $t=\sum_{i=1}^n t_i$ and $\tau_i=t_i/t$, which is valid, to
first order in $t$, for $t_i$ much shorter than the typical time
scale of the dynamics due to the Hamiltonian $A_i$. However, more
elaborate sequences of pulses (see Appendix \ref{err}) give
approximations to higher orders in $t$ \cite{sorn}. A recursive
application of the sequence (\ref{sHxyz}) determines an effective
evolution under the Hamiltonian $H_{\text{eff}}$. We point out
that, in this Hamiltonian, Eq. (\ref{Hxyz}), we can independently
control and change the values of the parameters $\tau_i$'s and
$\eta$, since they depend, respectively, on the free evolution
times $t_i$'s and on the pulse amplitude $b_s$. Consequently we
obtain an Hamiltonian $H_{\text{eff}}$ with a variable relative
strength of the spin-spin coupling in the $x$, $y$, and $z$
directions and a tunable transverse magnetic field. Notice that we
can also set $\tau_i=0$ for any desired $i$ or $\eta=0$. This is
achieved by simply choosing $t_i=0$ or switching off the external
field $\mathbf{b}_s(t)$. In this way various interesting quantum
spin models can be derived from Hamiltonian (\ref{sHxyz}). For
example for $\tau_2=\tau_3=0$ we obtain the Ising model, whereas
for $\tau_1=0$ ($\tau_2=0$) we obtain the $XY$ model in its usual
(rotated) basis.
\subsection{How to modify the interaction range and topology}

The spin Hamiltonian (\ref{Hzeta}) can be written as
\begin{equation}\label{Hpn}
\sum_{i>j}^N  J^z_{i,j} \sigma^z_i \sigma^z_j \equiv
H^z_{1}+H^z_{2}+\dots,
\end{equation}
where 
\begin{equation}
 H^z_{n} \equiv \sum_{i=1}^{N-n} J^z_{i,i+n} \sigma^z_i \sigma^z_{i+n} 
\end{equation}
represents the coupling between the $n$-th nearest neighbor spins.
In our system the
interaction between spins has a dipole-like nature, i.e. it
decreases with the third power of the inter-particle distance.
Consequently only the first few terms $H^z_{n}$ at the right hand
side of Eq. (\ref{Hpn}) play a significant role. In the following
we are going to outline a procedure to independently control and
modify, in a relatively simple way, the strength and the sign of
the relevant $H^z_{n}$ terms. In other words we can design the
interaction topology by enhancing or suppressing the coupling
between the $n$-th nearest neighbors.
This is achieved by iteratively applying to the system appropriate
sequences consisting of pulses alternated to periods of free
evolution. In our scheme each pulse affects simultaneously a
specific subset of spins in the array. Notice that if a particular
pulse sequence $S$ modifies the spin Hamiltonian $H^z$, we can extend the
same kind of coupling to the other directions by simply
performing, according to Eqs. (\ref{px}) and (\ref{py}), the
sequences $G_ySG_y^{\dagger}$ and $G_xSG_x^{\dagger}$.
Therefore, we restrict ourselves to the transformations affecting
the spin-spin coupling along the $z$ direction.
Furthermore, we are going to prove that
the number of pulses in each sequence does not depend on the number of
spins in the array, thus making our technique scalable with the
system size.

\begin{figure}
 \includegraphics{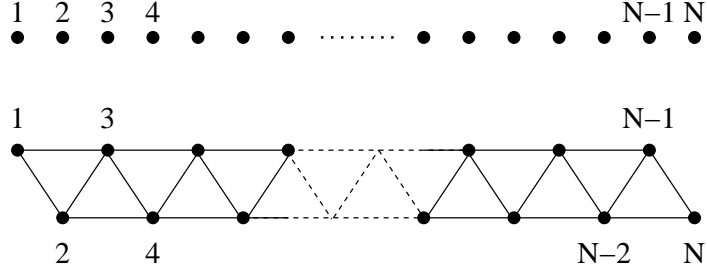}%
\caption{\label{2Darray}Schematic drawing of a linear chain of $N$
spins (upper part). When the first and second nearest neighbor
coupling strengths are made equal, the linear chain becomes equivalent
to a planar array (lower part).}
\end{figure}

As an example, we describe how to suppress the second nearest
neighbor interaction by making use
of three different transformations, defined as $\sigma^x_o$,
$\sigma^x_{c_1}$ and $\sigma_{c_2}^x$. Each of them can be
performed by a single multi-frequency pulse. The transformation
$\sigma^x_{o}$ consists in the simultaneous application of
$\sigma^x$ to all the spins in the odd sites of the array. The
transformations $\sigma^{x}_{c_1}$ and $\sigma^{x}_{c_2}$ flip,
instead, alternated couples of neighboring spins. In particular,
$\sigma_{c_1}^x$ affects the spin couples $\{1,2\},\{5,6\},\dots$,
whereas $\sigma^x_{c_2}$ affects the couples
$\{2,3\},\{6,7\},\dots$. We prove in Appendix \ref{seqs} that the
sequence
\begin{equation}
\sigma^x_{c_2}e^{-iH^z\frac{t}{2}}\sigma^x_o
 e^{-iH^z\frac{t}{2}}\sigma^x_{c_1}e^{-iH^z t}
\end{equation}
corresponds to the system evolution for a time $t$ under the
effective Hamiltonian $H_{\text{eff}}\simeq H^z_{1}+H^z_{3}$,
where the coupling of each spin with its second nearest neighbors
has been removed. Indeed, since the term $H^z_{3}$ is small, the
above sequence well approximates an effective NN Hamiltonian.  The
transformations $\sigma_o^x$, $\sigma^x_{c_1}$ and
$\sigma^x_{c_2}$ are the building blocks to construct other
sequences which realize different kinds of spin Hamiltonians. For
example, as described in Appendix \ref{seqs}, we can easily invert
the sign of $H^z$, thus switching from an anti-ferromagnetic to a
ferromagnetic interaction, or make the coupling strength between
first and second nearest neighbors equal. This last case
corresponds to an effective change in the array topology, since
the number of nearest neighbors passes from two (linear chain) to
four (see Fig. \ref{2Darray}).

In order to affect the coupling between neighboring spins of order
higher than three, we should apply simultaneously the
transformation $\sigma^x$ to selected subsets of three or more
spins along the chain. For example, as we show in Appendix
\ref{seqs}, with a seven pulse sequence we can suppress the
interaction between a spin and all its neighbors from the second
up to the sixth nearest neighbors. In such a way we improve our
approximation of a NN interacting spin chain.

\section{Fidelity} \label{fid}
In this section we discuss the performances of the scheme, based
on the detailed analysis reported in Appendix \ref{err}. 
We emphasize that our treatment focuses on the limitations
due to the mapping of the system of trapped electrons into the
desired target system of interacting spins.
Hence, most experimental imperfections are not considered here.

As described in the previous section, to derive the effective spin
Hamiltonian we make use of the approximate identity (\ref{tro}) or
of more sophisticated approximations \cite{sorn}.
As a consequence we
introduce an error \cite{chuang}
\begin{equation}
  \mathcal{E} \equiv \| U-U' \|  \equiv
  \text{max}_{|\psi\rangle:||\psi\rangle|=1}|(U-U')|\psi\rangle|,
\end{equation}
which measures the distance between the desired evolution $U$ and
the approximated evolution $U'$.
For instance, in our case the target unitary operator $U=\exp(-iH_{\text{eff}}t)$,
with the effective spin Hamiltonian $H_{\text{eff}}$ of the kind of Eq. (\ref{Hxyz}), is approximated to the fourth order in $t$
by the sequence $S$, Eq. (\ref{tro4}) \cite{sorn}.
By using some algebra (see Appendix~\ref{err}), we can bound the
error associated to the application of a single sequence $S$ with
\begin{equation}\label{Eerr}
\mathcal{E}_S \leq (J^z t)^5 f(N) ,
\end{equation}
where $J^z \equiv J^z_{i,i+1}$ is the NN coupling strength and $f(N)$ is, in good approximation,
an increasing linear
function of the number $N$ of electrons in the array.
The exact
form of $f(N)$ depends on the specific target spin Hamiltonian.
From this result, we see that the error is small whenever the time
evolution is much shorter than the
flipping time, i.e. $t \ll \pi/J^z$.
We prove in Appendix~\ref{err} that, if we iterate $m$ times the
sequence $S$ the total error is $\mathcal{E} \leq m
\mathcal{E}_S$.
Therefore, if we want
to simulate the system evolution for a given time $T=mt$, to keep the
accuracy high we should apply the same sequence $S$ $m$ times
\begin{equation}
 \mathcal{E} \leq \frac{(J^z T)^5 f(N)}{m^4} .
\end{equation}
For a given simulation time $T$ and coupling strength $J^z$, the error $\mathcal{E}$ decreases with the number of iterations $m$ and, therefore, with the total number of pulses.

In appendix \ref{err} we provide the explicit
expression of $f(N)$ for the $XY$ and NN Ising models. Consequently we are able
to estimate the upper bound of $\mathcal{E}$ in both cases.
In our analysis, we also take into account the error introduced by the
derivation of the effective spin-spin coupling \cite{spinchain}. In particular, the error
$\mathcal{E}_c$ due to the canonical transformation (\ref{S})
satisfies the relation
\begin{equation}\label{ecan}
\mathcal{E}_c \leq N \left( \bar{k}+\frac{1}{2} \right) \varepsilon^2,
\end{equation}
where $\bar{k}$ is the mean axial oscillator excitation number.
We consider an array of $50$ electrons with inter-particle
distance of 100 $\mu$m, $\omega_{s}/2 \pi = 100$ GHz,
$\omega_z/2 \pi = 160$ MHz, and a magnetic gradient $b \simeq 200$ T/m.
With these parameters we obtain, according to Eq. (\ref{Jz}),
a NN coupling constant $J^z = 10$ Hz.
We also assume that the spin frequencies of neighboring electrons differ of about 2 MHz,
each pulse has a duration of the order of $\mu$s and the axial motion is cooled to the ground state.
In order to simulate the $XY$ model (NN Ising model)
for a time $T = 1$ s with fidelity of $99\%$, we need to iterate
the specific sequence $S$ about 100 (50) times.
In particular the simulation of the $XY$ model
requires about 3000 pulses, whereas the Ising model
with NN coupling requires about 2000 pulses.
Notice that the Ising model with dipole-like
coupling requires no pulse sequence, since it is obtained, directly, by applying the field (\ref{bs}).
Therefore, in this case we only take into account, as a source of error,
the thermal excitation of
the axial oscillator which, according to Eq. (\ref{ecan}), 
is of the order of $10^{-3}$.

\section{Conclusions} \label{concl}

In this paper we have proposed a scalable technique for easily controlling
and adjusting the effective Hamiltonian of a system of interacting spins.
The underlying physical system consists of an array of trapped electrons in micro-Penning traps.
The electron spin is prepared and manipulated with an external resonant magnetic field.
These spin operations, applied to all the particles or subsets of them, are alternated to periods of free evolution in a fashion similar to NMR refocusing schemes. To selectively address the electrons in the array, it is necessary to introduce a detuning between the characteristic spin frequencies by means of a magnetic gradient. In particular, we have shown that, in the case of a
system without spin frequency differentiation, a two pulse sequence
permits to reduce or even cancel the effect on the spin dynamics of the uniform magnetic field, without affecting the overall trap stability.
This is potentially useful for the observation of quantum phase transitions \cite{sachdev}, where it is important to modulate the ratio between the external magnetic field and the spin-spin coupling strength. In the case of a system with
spin frequency differentiation, we have proved that with a
repeated application of appropriate pulse sequences we can modify and
control the interaction terms in the effective spin Hamiltonian. As a
result a wide range of spin Hamiltonians can be obtained, such as
the Ising model and the $XY$ models. Moreover, specific
pulse sequences allow to control the sign and strength of the
coupling between the $k$-th nearest neighbors for any
significant value of $k$ (first, second, \ldots, nearest neighbors).
As an example, we provide a prescription to obtain an Hamiltonian
with substantially only NN coupling starting from a dipole-like interaction.
In our scheme the number of pulses in each sequence is relatively
small and does not depend on the number of spins in the array.
We derive an analytical formula to estimate the fidelity of our method 
for simulating the effective spin Hamiltonian,
as a function of the coupling strength, the simulation time 
and the number of particles.
Our estimates show that it is feasible to simulate the Ising, with NN coupling, and the $XY$ model
with fidelity of $99\%$ for a system of $50$ electrons with 
a coupling strength $J^z = 10$ Hz.
Of course, the evaluation of the performances of a real experiment would
require a closer analysis of all the possible sources
of errors and decoherence. This is, however, beyond the scope of the 
present work.
\appendix

\section {} \label{seqs}

In this appendix, we are going to prove that spin flip operations,
applied to subsets of particles in the array, result in an
effective sign change in the interaction between neighbors of
arbitrary order. The starting point is represented by the relation
\begin{equation}\label{xzx}
\sigma^x_i \sigma^z_i \sigma^x_i=-\sigma^z_i,
\end{equation}
which reverts the sign of the $i$-th spin operator. We define the
following operators
\begin{eqnarray}
\sigma^x_o &\equiv& \bigotimes_{i=1}^{N/2} \sigma^x_{2i-1},\label{single}  \\
\sigma^x_{c_k} &\equiv& \bigotimes_{i\in c_k} \sigma^x_i
\sigma^x_{i+1} \qquad  \text{with} \
c_k=\{k,k+4,k+8,\dots\} \ \text{for} \ k=1,2, \label{pair}\\
\sigma^x_{\mathcal{T}_k} &\equiv& \bigotimes_{i\in\mathcal{T}_k}
\sigma^x_i \sigma^x_{i+1} \sigma^x_{i+2} \qquad \text{with} \
\mathcal{T}_k=\{k,k+6,k+12,\dots\} \ \text{for} \ k=1,2,3, \label{tern}\\
\sigma^x_{\mathcal{Q}_k}&\equiv&\bigotimes_{i\in\mathcal{Q}_k}
\sigma^x_i \sigma^x_{i+1} \sigma^x_{i+2} \sigma^x_{i+3} \quad
\text{with} \ \mathcal{Q}_k=\{k,k+8,k+16,\dots\} \ \text{for} \
k=1,2,3,4,\label{quat}
\end{eqnarray}
that affect simultaneously different subsets of spins. Moreover,
we observe that
\begin{equation}
(\sigma^x_o)^2=(\sigma^x_{c_k})^2=(\sigma^x_{\mathcal{T}_k})^2 =
(\sigma^x_{\mathcal{Q}_k})^2 = \openone
\end{equation}
for any possible value of $k$, so that we can use the identity
\begin{equation}\label{ABA}
Ae^BA=\exp{(ABA)},
\end{equation}
which holds true for any pair of operators $A$ and $B$, whenever
$A^2=\openone$.

From the relation (\ref{xzx}) and the definition of $\sigma^x_o$
it follows that
\begin{equation}
\sigma^x_o \sigma^z_i \sigma^z_j \sigma^x_o=(-1)^{i+j}\sigma^z_i
\sigma^z_j,
\end{equation}
which amounts to a sign change in the interaction between spins
with different parity. Consequently, given the Hamiltonian $H^z$
of Eq. (\ref{Hzeta}), the transformation $\sigma^x_o H^z
\sigma^x_o$ inverts the coupling between neighbors of odd orders
\begin{equation}\label{oHo}
\sigma_{o}^x e^{-iH^z t} \sigma_{o}^x =e^{-i\sigma_o^x
H^z\sigma_o^xt}=\exp{\left[-i\sum_{k=1}^N (-1)^k H^z_k t\right]}.
\end{equation}
This property allows us to make equal in strength the coupling
between first and second nearest neighbors
\begin{equation}\label{seq}
\sigma_{o}^x e^{-iH^z \frac{7}{9}t} \sigma_{o}^x e^{-iH^z t}\simeq
\exp{\left[-iJ'\sum_{j=1}^{N-2}\sigma^z_j(\sigma^z_{j+1}+\sigma^z_{j+2})t\right]},
\end{equation}
with $J'\equiv (2/9)J^z_{i,i+1}$.

With a three-pulse sequence
\begin{equation}\label{Hzc1c2}
\sigma^x_{c_2}e^{-iH^z\frac{t}{2}}\sigma^x_o e^{-iH^z
\frac{t}{2}}\sigma^x_{c_1} =\exp{\left[-i\sum_{k=1}^{N/2} (-1)^k
H^z_{2k}t\right]},
\end{equation}
we remove the coupling between odd order neighbors and
alternatively change the sign of the coupling in $H^z$ between
even order neighbors. To prove Eq. (\ref{Hzc1c2}) we use
the identity $\sigma^x_o=\sigma^x_{c_2}\sigma^x_{c_1}$ and the
commutation relation
$[\sigma^x_{c_2}H^z\sigma^x_{c_2},\sigma^x_{c_1}H^z\sigma^x_{c_1}]=0$
in order to obtain
\begin{equation}
\sigma^x_{c_2}e^{-iH^z\frac{t}{2}}\sigma^x_o e^{-iH^z
\frac{t}{2}}\sigma^x_{c_1} =\sigma^x_{c_2}e^{-iH^z
\frac{t}{2}}\sigma^x_{c_2} \sigma^x_{c_1}e^{-iH^z
\frac{t}{2}}\sigma^x_{c_1}=e^{-i(\sigma^x_{c_2}H^z\sigma^x_{c_2}+\sigma^x_{c_1}H^z\sigma^x_{c_1})\frac{t}{2}}.
\end{equation}
The transformation $\sigma^x_{c_k}H^z\sigma^x_{c_k}$ selectively
changes the sign in $H^z$ to the operators $\sigma^z_j$ and
$\sigma^z_{j+1}$ with $j \in c_k$, according to Eq. (\ref{pair}).
Consequently we have
\begin{eqnarray}\label{c1Hzc1}
\sigma^x_{c_1}H^z\sigma^x_{c_1}&=&\sum_{i=1}^{N-1}(-1)^{i+1}
J^z_{i,i+1} \sigma^z_i\sigma^z_{i+1}-\sum_{i=1}^{N-2} J^z_{i,i+2}
\sigma^z_i\sigma^z_{i+2}+\sum_{i=1}^{N-3}(-1)^{i} J^z_{i,i+3}
\sigma^z_i\sigma^z_{i+3}+\dots, \\
\label{c2Hzc2}
\sigma^x_{c_2}H^z\sigma^x_{c_2}&=&\sum_{i=1}^{N-1}(-1)^{i}
J^z_{i,i+1} \sigma^z_i\sigma^z_{i+1}-\sum_{i=1}^{N-2} J^z_{i,i+2}
\sigma^z_i\sigma^z_{i+2}+\sum_{i=1}^{N-3}(-1)^{i+1} J^z_{i,i+3}
\sigma^z_i\sigma^z_{i+3}+\dots.
\end{eqnarray}
Hence, to demonstrate Eq. (\ref{Hzc1c2}), we sum Eq.
(\ref{c1Hzc1}) and Eq. (\ref{c2Hzc2}) obtaining
\begin{equation}
\sigma^x_{c_2}H^z\sigma^x_{c_2}+\sigma^x_{c_1}H^z\sigma^x_{c_1}=-2H^z_2+2H^z_4+\dots=2\sum_{k=1}^{N/2}(-1)^k
H^z_{2k}
\end{equation}
Notice that in the sum the coupling between nearest neighbors of
odd orders cancels out.

By combining the sequences (\ref{oHo}) and (\ref{Hzc1c2}), we can
invert the sign of the coupling up to third nearest neighbors
\begin{equation} \label{sinv}
(\sigma^x_{c_2}e^{-iH^z t}\sigma^x_o
 e^{-iH^z t}\sigma^x_{c_1}) (\sigma_{o}^x e^{-iH^zt}
 \sigma_{o}^x)\simeq \exp{\left[-i(-H^z_1-H^z_2-H^z_3)t\right]},
\end{equation}
thus turning a ferromagnetic interaction into an anti-ferromagnetic
one and viceversa.
Another consequence of Eq. (\ref{Hzc1c2}) is
\begin{equation}
\sigma^x_{c_2}e^{-iH^z\frac{t}{2}}\sigma^x_o
 e^{-iH^z\frac{t}{2}}\sigma^x_{c_1}e^{-iH^z
 t}\simeq\exp{\left[-i(H^z_1+H^z_3)t\right]},
\end{equation}
where the coupling between second nearest neighbors has been
removed.

The approach described so far can be extended in order to cancel
coupling terms of higher order. For example, to remove both the
second and third nearest neighbor couplings in $H^z$ we make use
of the transformations defined in Eq. (\ref{tern}). They
simultaneously affect sets of three nearest neighbors in alternate
succession. With arguments similar to those used for verifying Eq.
(\ref{Hzc1c2}), we can demonstrate the following identity
\begin{equation}\label{HzT}
(\sigma^x_{\mathcal{T}_3}e^{-iH^z\frac{t}{3}}\sigma^x_{\mathcal{T}_3})(\sigma^x_{\mathcal{T}_2}e^{-iH^z\frac{t}{3}}\sigma^x_{\mathcal{T}_2})
(\sigma^x_{\mathcal{T}_1}e^{-iH^z\frac{t}{3}}\sigma^x_{\mathcal{T}_1})=
e^{-i (H^z_{1} - H^z_{2} - 3 H^z_{3} - H^z_{4} + \dots )t/3}.
\end{equation}
By combining Eq. (\ref{Hzc1c2}) and Eq. (\ref{HzT}) we prove that
the sequence
\begin{equation}\label{canc23}
(\sigma^x_{c_2} e^{-iH^z\frac{t}{3}} \sigma^x_{o}
e^{-iH^z\frac{t}{3}} \sigma^x_{c_1})(\sigma^x_{\mathcal{T}_3}
e^{-iH^z\frac{t}{3}} \sigma^x_{\mathcal{T}_3})
(\sigma^x_{\mathcal{T}_2} e^{-iH^z\frac{t}{3}}
\sigma^x_{\mathcal{T}_2})(\sigma^x_{\mathcal{T}_1}
e^{-iH^z\frac{t}{3}}\sigma^x_{T_1}) e^{-iH^z t}
\end{equation}
corresponds to the evolution for a time $(4/3)t$ under the
Hamiltonian $(H^z_{1}+H^z_{4}+\dots)$, where the coupling between
second and third nearest neighbors has been removed. The
implementation of this sequence requires six pulses, since each
couple of consecutive transformations in Eq. (\ref{canc23}) is
equivalent to a single transformation affecting simultaneously a
specific subset of spins in the array.

It is worth to point out that with a seven-pulse sequence we can
approximate the NN model in a very accurate way, i.e. we can
suppress the interaction between a spin and all its neighbors from
the second up to the sixth nearest neighbors. This is achieved by
using the four transformations defined in Eq. (\ref{quat}), that
simultaneously affect alternated sets of four nearest neighbors.
It can be proved that
\begin{equation}\label{HzQ}
(\sigma^x_{\mathcal{Q}_4}e^{-iH^z\frac{t}{2}}\sigma^x_{\mathcal{Q}_4})(\sigma^x_{\mathcal{Q}_3}e^{-iH^z\frac{t}{2}}\sigma^x_{\mathcal{Q}_3})
(\sigma^x_{\mathcal{Q}_2}e^{-iH^z\frac{t}{2}}\sigma^x_{\mathcal{Q}_2})(\sigma^x_{\mathcal{Q}_1}e^{-iH^z\frac{t}{2}}\sigma^x_{\mathcal{Q}_1})=
e^{-i\left(H^z_{1}-H^z_{3}-2H^z_{4}-H^z_{5}+H^z_{7}+\dots\right)t}.
\end{equation}
From Eq. (\ref{Hzc1c2}) and Eq. (\ref{HzQ}) we have that the
sequence
\begin{equation}\label{canc26}
(\sigma^x_{c_2} e^{-iH^z\frac{t}{2}} \sigma^x_{o}
e^{-iH^z\frac{t}{2}}
\sigma^x_{c_1})(\sigma^x_{\mathcal{Q}_4}e^{-iH^z\frac{t}{2}}\sigma^x_{\mathcal{Q}_4})(\sigma^x_{\mathcal{Q}_3}e^{-iH^z\frac{t}{2}}\sigma^x_{\mathcal{Q}_3})
(\sigma^x_{\mathcal{Q}_2}e^{-iH^z\frac{t}{2}}\sigma^x_{\mathcal{Q}_2})(\sigma^x_{\mathcal{Q}_1}e^{-iH^z\frac{t}{2}}\sigma^x_{\mathcal{Q}_1})
e^{-iH^z t}
\end{equation}
corresponds to the evolution for a time $2t$ under the Hamiltonian
$(H^z_{1}+H^z_{7}+\dots)$, where the coupling between nearest
neighbors from the second up to the sixth order has been removed.
The implementation of the above sequence requires just seven pulses, 
since each couple of
consecutive transformations in Eq. (\ref{canc26}) is equivalent to
a single transformation affecting simultaneously a specific subset
of spins in the array.

\section {} \label{err}

The sequence of unitary operators to the left hand side of
relation (\ref{tro}) approximates the evolution under the
target Hamiltonian
\begin{equation} \label{B1}
  H = \sum_{i=1}^n \tau_i A_i
\end{equation}
to first order in $t$.
However, more elaborate combinations of unitary operators
provide better approximations.
For example the sequence \cite{sorn}
\begin{equation}\label{tro4}
S\equiv\bar{S}_1 S_1 \bar{S}_1  S_{-2} \bar{S}_1 \bar{S}_1
\bar{S}_1 \bar{S}_1 S_1 \bar{S}_1 S_1 S_1 S_1 S_1 \bar{S}_{-2} S_1
\bar{S}_1 S_1
\end{equation}
with
\begin{equation}
S_k \equiv e^{-i\frac{k}{12}t_1 A_1} e^{-i\frac{k}{12}t_2
A_2}\dots e^{-i\frac{k}{12}t_n A_n}
\end{equation}
and
\begin{equation}
\bar{S}_k \equiv e^{-i\frac{k}{12}t_n A_n} \dots
e^{-i\frac{k}{12}t_2 A_2} e^{-i\frac{k}{12}t_1 A_1},
\end{equation}
approximates the unitary operator $e^{-i H
t}$, with $t \equiv \sum_{i=1}^n t_i$ and $\tau_i \equiv t_i/t$, to
the fourth order in $t$.

The error introduced by
approximating the unitary operator $U$ with the unitary operator $U'$ can be
measured by the quantity \cite{chuang}
\begin{equation}
 \mathcal{E} \equiv \|U-U'\|  \equiv
\text{max}_{|\psi\rangle:||\psi\rangle|=1}|(U-U')|\psi\rangle|.
\end{equation}
Now if we want to approximate the evolution under the Hamiltonian
Eq.~(\ref{B1}) for a time $T=m t$ we can apply $m$
times the sequence (\ref{tro4}).
From the generalization of the inequality
\begin{equation}
  \|A B - C D \|\leq \|A - C\|+\|B - D \| ,
\end{equation}
verified for any
unitary operator $A$, $B$, $C$, $D$, we have that the
error $\mathcal{E}$ of our approximation satisfies the inequality
$\mathcal{E}\leq\sum_{i=1}^m \mathcal{E}_i$, where $\mathcal{E}_i$
is the error introduced by the $i$-th application of the
sequence (\ref{tro4}).
More specifically we evaluate the error
\begin{equation}\label{Serr}
\mathcal{E}_i \equiv \|e^{-i\left(\sum_{i=1}^n \tau_i A_i\right)t}-S \| ,
\end{equation}
when the operators $A_i$'s are typically of the kind
$\sum_{i>j} J_{i,j}^z \sigma_i^k \sigma_j^k$,
with $k = x,y,z$.
We  explicitly expand each operator to the right hand side and find
that their difference is proportional to $t^5$, because the sequence $S$
approximates the desired unitary evolution $\exp(-i \sum_{i=1}^n \tau_i A_i t)$
to the fourth order in $t$.
Moreover, we make use of the inequality
\begin{equation}
\|\alpha A+\beta B \| \leq |\alpha|\|A\|+|\beta|\|B\|,
\end{equation}
which holds true  for any pair of operators
$A$ and $B$ and complex numbers $\alpha$, $\beta$.
Finally, we observe that $\|C\|=1$ if $C$ is any product of Pauli operators
$\sigma^k_i$.
This approach lead us to the following estimate for the error, defined in
Eq.~(\ref{Serr}),
\begin{equation}
\mathcal{E}_i \leq (J^z t)^5 f(N) ,
\end{equation}
where $J^z \equiv J^z_{i,i+1}$ is the nearest neighbor coupling strength
and $f(N)$ is, in good approximation, an increasing linear
function of the number $N$ of electrons in the array, depending on the specific form of the spin
Hamiltonian.

When we apply $m$ times the sequence $S$, from the previous discussion it
follows that $\mathcal{E} \leq m(J^z t)^5 f(N)$.
Now if we indicate with $T=m t$ the total simulation time, we
obtain
\begin{equation}
\mathcal{E}\leq \frac{ (J^z T)^5 f(N)}{m^4}
\end{equation}
or, equivalently,
\begin{equation}\label{sqrt4}
m \leq  \sqrt[4]{\frac{(J^z T)^5 f(N)}{\mathcal{E}}}.
\end{equation}
Equation (\ref{sqrt4}) gives an upper bound to the number of
iterations required to mimic the desired evolution with an error
$\mathcal{E}$. For example, to approximate the $XY$
model in the rotated basis, we choose
\begin{eqnarray}
A_1 &=& \frac{\eta}{2} \sum_{i=1}^N \sigma_i^x+\sum_{i>j}^N J_{i,j}^z \sigma_i^z \sigma_j^z , \\
A_2 &=& \frac{\eta}{2} \sum_{i=1}^N \sigma_i^x+\sum_{i>j}^N J_{i,j}^z \sigma_i^y \sigma_j^y.
\end{eqnarray}
In this case, the outlined approach gives
for $N>5$, $f(N) \simeq  0.25 N-0.85$.
The simulation of the NN Ising model is achieved with
\begin{eqnarray}
A_1 &=& \frac{\eta}{2} \sum_{i=1}^N \sigma_i^x +
        \sum_{i>j}^N J_{i,j}^z \sigma_i^z \sigma_j^z , \\
A_2 &=&\frac{\eta}{2} \sum_{i=1}^N \sigma_i^x -\sum_{i=1}^{N-2} J_{i,i+2}^z \sigma_i^z \sigma_{i+2}^z,
\end{eqnarray}
and, for $N>5$, $f(N) \simeq  0.015 N-0.035$.
\begin{acknowledgments}
This research was supported by the European Commission through the
Specific Targeted Research Project \emph{QUELE}, the Integrated Project
FET/QIPC \emph{SCALA}, and the Research Training Network \emph{CONQUEST}.
\end{acknowledgments}

\end{document}